\documentclass[12pt]{iopart}
\usepackage{iopams}
\usepackage[caption=false]{subfig}
\usepackage{graphicx}
\usepackage[]{hyperref}

\begin{document}

\title[Energy spread of ultracold electron bunches extracted from a laser cooled gas]{Energy spread of ultracold electron bunches extracted from a laser cooled gas}

\author{J.G.H. Franssen$^{1,2}$, J.M. Kromwijk$^{1}$, E.J.D. Vredenbregt$^{1,2}$, O.J. Luiten$^{1,2,3}$}
\address{$^1$Department of Applied Physics, Eindhoven University of Technology, P.O. Box 513, 5600 MB Eindhoven, The Netherlands}
\address{$^2$Institute for Complex Molecular Systems, Eindhoven University of Technology, P.O. Box 513, 5600 MB Eindhoven, The Netherlands}
\ead{$^3$\mailto{o.j.luiten@tue.nl}}

\begin{abstract}

Ultrashort and ultracold electron bunches created by near-threshold femtosecond photoionization of a laser-cooled gas hold great promise for single-shot ultrafast diffraction experiments. In previous publications the transverse beam quality and the bunch length have been determined. Here the longitudinal energy spread of the generated bunches is measured for the first time, using a specially developed Wien filter. The Wien filter has been calibrated by determining the average deflection of the electron bunch as a function of magnetic field. The measured relative energy spread $\frac{\sigma_{U}}{U}=0.64\pm0.09\%$ agrees well with the theoretical model which states that it is governed by the width of the ionization laser and the acceleration length.

\end{abstract}

\vspace{2pc}
\submitto{\jpb}

\maketitle


\section{Introduction}

One of the dreams in physics, chemistry and materials science is to be able to monitor the dynamical behavior of matter with atomic spatial and temporal resolution, i.e. $0.1$~nm and $100$~fs, and thus enable the investigation\cite{Sciaini2011a,King2005} of phase transitions, chemical reactions and conformational changes at the most fundamental level.
One of the techniques which has emerged recently and is developing at a rapid pace is ultrafast electron diffraction\cite{Sciaini2011a} (UED). This technique uses femtosecond laser pulses to excite a dynamical process in a crystalline material, which is subsequently probed by a synchronized, ultrashort electron bunch.

Measurements of transverse beam quality have shown that electron pulses as cold as $10$~K can be produced using femtosecond photoionization\cite{Engelen2013,Engelen2014}. It has also been shown that the ultracold electron source is capable of producing high quality diffraction patterns\cite{VanMourik2014a,Speirs2015a}.
The pulse length of the ultracold electron bunches extracted from a laser cooled gas was measured using a RF deflecting cavity\cite{Franssen2017} and resulted in pulse lengths of $\sim 20~$ps for femtosecond laser ionization. Similarly electron pulse lengths of $130~$ps were measured using a high voltage deflector when using a two-color multi-photon excitation process\cite{PhysRevA.95.053408}.

 These measured pulse lengths were limited by the energy spread of the electron beam. The energy spread causes degradation in temporal resolution by the lengthening of the electron bunches. This effect can be cancelled by compressing the electron pulses using established RF techniques\cite{VanOudheusden2010a,Pasmans2013}. The ultimate temporal resolution is governed by the longitudinal emittance which is determined by the uncorrelated pulse length and energy spread. The temporal resolution in UED experiments can be further improved by cancelling the arrival time jitter induced by RF phase fluctuations\cite{Franssen2017a} in bunch compression cavities.

The energy spread of the electron bunches extracted from the ultracold electron source has not yet been measured directly. This article presents the first experiments using a specially designed Wien filter. The device can eventually be used to investigate the effects of space charge forces in picosecond electron bunches. Simultaneous measurement of both the energy spread and pulse length will allow us to determine the longitudinal phase space distribution and thus the ultimately achievable temporal resolution in UED experiments.


\section{The ultra cold electron source}

\begin{figure}[htb!]
\centering
\includegraphics[width=14cm]{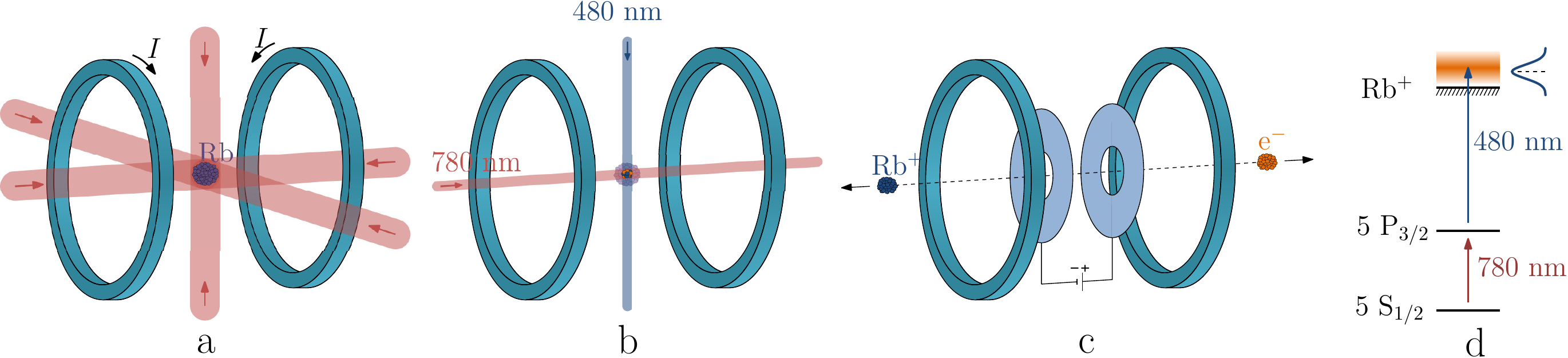}	
\caption{\label{mot}Schematic representation of the electron bunch production sequence. The six red beams in figure~(a) represent the cooling laser beams and the two blue coils the magnetic field coils in anti-Helmholtz configuration. In the figure~(b) the red beam indicates the excitation laser and the blue beam the ionization laser. Figure~(c) illustrates the acceleration process in a static electric field. Figure~(d) illustrates the two step ionization scheme. The atoms are first excited from the $5S_{1/2}F=3$ state to the $5P_{3/2}F=4$ state, from which they are ionized using a $480~$nm photon.}
\end{figure}

The ultracold electron bunches are created by near-threshold femtosecond photoionization of laser cooled and trapped 85-rubidium atoms, as is illustrated in \Fref{mot}. First, \Fref{mot}a, rubidium atoms are trapped and cooled in a magneto optical trap (MOT). Second, \Fref{mot}b, the trapping laser is temporarily switched off for $1~\mu$s, so that the atoms relax back to their ground state. After the atoms have relaxed to the $5S_{\frac{1}{2}}F=3$ state, a $780$~nm excitation laser beam is switched on, creating a small cylinder (radius of $25~\mu$m) of excited atoms in the $5P_{\frac{3}{2}} F=4$ state. Subsequently a small volume of rubidium atoms is ionized by a pulsed $480$~nm femtosecond ionization laser beam, \Fref{mot}d, intersecting the excitation laser beam at right angles, resulting in a cloud of cold electrons and ions. The shape and size of the ionization volume can be controlled by the overlap of the excitation and ionization laser beams\cite{PhysRevLett.95.164801,McCulloch2011}. Finally, \Fref{mot}c, the electrons are extracted by a static electric field\cite{Taban2008}. 

\begin{figure*}[htb!]
\centering
\includegraphics[width=14cm]{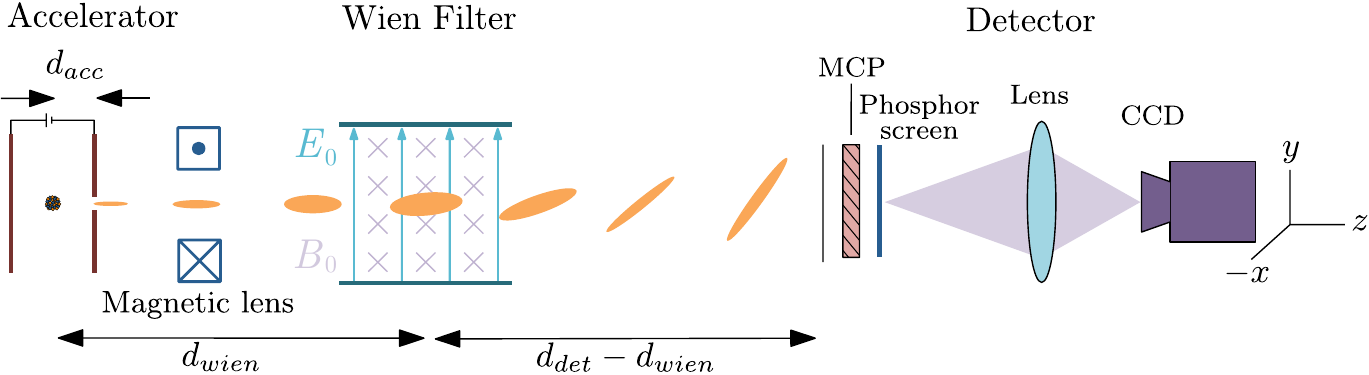}
\caption{\label{beamline}Schematic representation of the beam line, consisting of an electrostatic accelerator, a magnetic solenoid lens, the Wien filter and a MCP with a phosphor screen. The light emanating from the phosphor screen is imaged onto a CCD camera.}
\end{figure*}

\Fref{beamline} shows a schematic representation of the electron beam line. First the electron beam passes through a magnetic solenoid lens which is used to focus the beam onto the detector. The Wien filter is positioned a distance $d_{wien}=0.68~$m from the source. The electrons are detected by a micro channel plate in combination with a phosphor screen, positioned at a distance $d_{det}=1.9~$m from the source.

The longitudinal energy spread, in absence of space charge forces, is to a good approximation governed by the width of the ionization laser beam in the direction of the acceleration field\cite{Franssen2017,Reijnders2009}. The relative energy spread is given by

\begin{equation}\frac{\sigma_{U}}{U}=\frac{\sigma_{\rm ion}}{d_{acc}}\label{relenergylasersize},\end{equation}

with $\sigma_{\rm ion}$ the rms size of the ionization laser profile in the acceleration ($\hat{z}$) direction, $\sigma_{U}$ the rms energy spread, $U=eE_{acc}d_{acc}$ the average bunch energy, $e$ the elementary charge, $E_{acc}$ the acceleration field and $d_{acc}=13.5~$mm the effective acceleration length\cite{Taban2008,Reijnders2009}. The initial energy spread due to thermal motion is negligible compared to the energy spread due to the finite ionization laser beam size\cite{Franssen2017} which means that we can neglect the Boersch effect\cite{Boersch1954}.

\section{Wien filter}

A Wien filter is an electro-magnetic element that separates charged particles by their velocity\cite{Wienfilterart}. An ideal Wien filter consists of a static and uniform magnetic field, with magnitude $B_{0}$, perpendicular to a static and uniform electric field, with magnitude $E_{0}$. Both fields should be perpendicular to the direction of the electron beam that is passing through. A schematic representation of an ideal Wien filter is shown in \Fref{beamline}.

When an electron is injected into such an ideal field, it will gain a transverse momentum with a magnitude that is dependent on the particle velocity. The particles will exit the Wien filter with a transverse momentum distribution that is correlated to the initial longitudinal momentum distribution. This transverse momentum distribution will result in a spread on a detector screen which therefore maps the longitudinal momentum distribution on the screen.

For an electron beam moving in the $\hat{z}$ direction the longitudinal velocity $v_{z}$ will be much greater than the transverse velocities: $v_{z}>>v_{x},v_{y}$. In a Wien filter with a uniform electric field $\vec{E}=E_{0}~\hat{y}$ and a uniform magnetic field $\vec{B}= B_{0}~\hat{x}$, that both extend over a length $L_{w}$ in the $\hat{z}$ direction, a particle with velocity $\vec{v}=v_{z}~\hat{z}$ is deflected by an angle

\begin{equation}\theta_{y}=\theta_{c} \left(1-\frac{E_{0}}{v_{z} B_{0}}\right).\label{defllor}\end{equation}

With $\theta_{c}\equiv \omega_{c}t_{w}$ the cyclotron angle, $\omega_{c}= \frac{eB_{0}}{m}$ the cyclotron frequency and $t_{w}=\frac{L_{w}}{v_{z}}$ the time spent inside the Wien filter. \Eref{defllor} shows us that a particle with a velocity

\begin{equation}\vec{v}_{c}=\frac{E_{0}}{B_{0}}\hat{z}\label{Wiencriterium}\end{equation}

is not deflected.

\begin{figure*}[htb!]
\centering
\includegraphics[width=12cm]{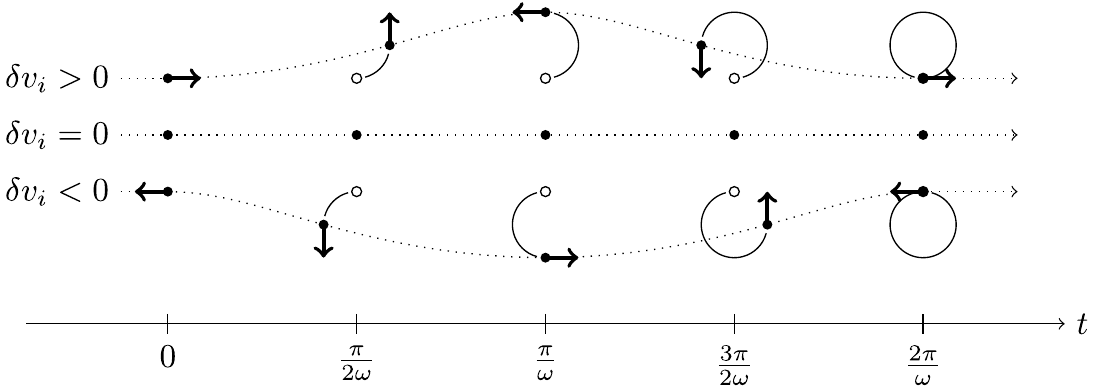}
\caption{\label{Wiencyclmotion}Illustration of the longitudinal and transverse momentum exchange, described by \Eref{wienexchange1} and \Eref{wienexchange2}, as a function of cyclotron angle $\theta_{c}$ for different particles.}\end{figure*}

A particle with a velocity $\vec{v}(t)=\vec{v}_{c}+\delta\vec{v}(t)$ entering the Wien filter at time $t=0$ will be deflected by gaining additional longitudinal and transverse velocities described by
\numparts  
\begin{eqnarray}
\label{wienexchange1}
\delta v_{y}(t)=\delta v_{z}(0) \sin(\omega_{c}t),\\
\label{wienexchange2}
\delta v_{z}(t)=\delta v_{z}(0) \cos(\omega_{c}t),
\end{eqnarray}
\endnumparts

with $t$ the time spent inside the Wien filter. At one quarter of a cyclotron orbit, $\theta_{c}=\frac{\pi}{2}$, the longitudinal velocity $\delta v_{z}(0)$ is fully converted into transverse velocity; this is illustrated in \Fref{Wiencyclmotion}. 

The cyclotron angle $\theta_{c}$ can be calibrated by measuring the change in average deflection angle $\Delta \theta_{y}$ of an electron pulse when varying the magnetic field by an amount $\Delta B$. These two quantities are related by

\begin{equation}
\Delta \theta_{y} = \theta_{c}~\frac{\Delta B}{B_{0}},\label{caldeflec}
\end{equation}

which is easily verified by expanding \Eref{defllor} for small $\frac{\Delta B}{B_{0}}$ around $\theta_{y}=0$. Knowing the cyclotron angle $\theta_{c}$, we can calculate the relative energy spread by measuring the rms streak length $\sigma_{y}$ on a screen. The rms streak length is given by

\begin{equation}
\sigma_{y}^{2}=\sigma_{\rm off}^{2}+ \left(\frac{d_{det}-d_{wien}}{2} \frac{\sigma_{U}}{U}\sin(\theta_{c})\right)^{2}\label{energyfit},
\end{equation}

with $\sigma_{\rm off}$ the rms transverse beam size when the Wien filter is turned off.

\subsection{Wien filter design}

\begin{figure}[htbp]
\centering
\includegraphics[width=10cm]{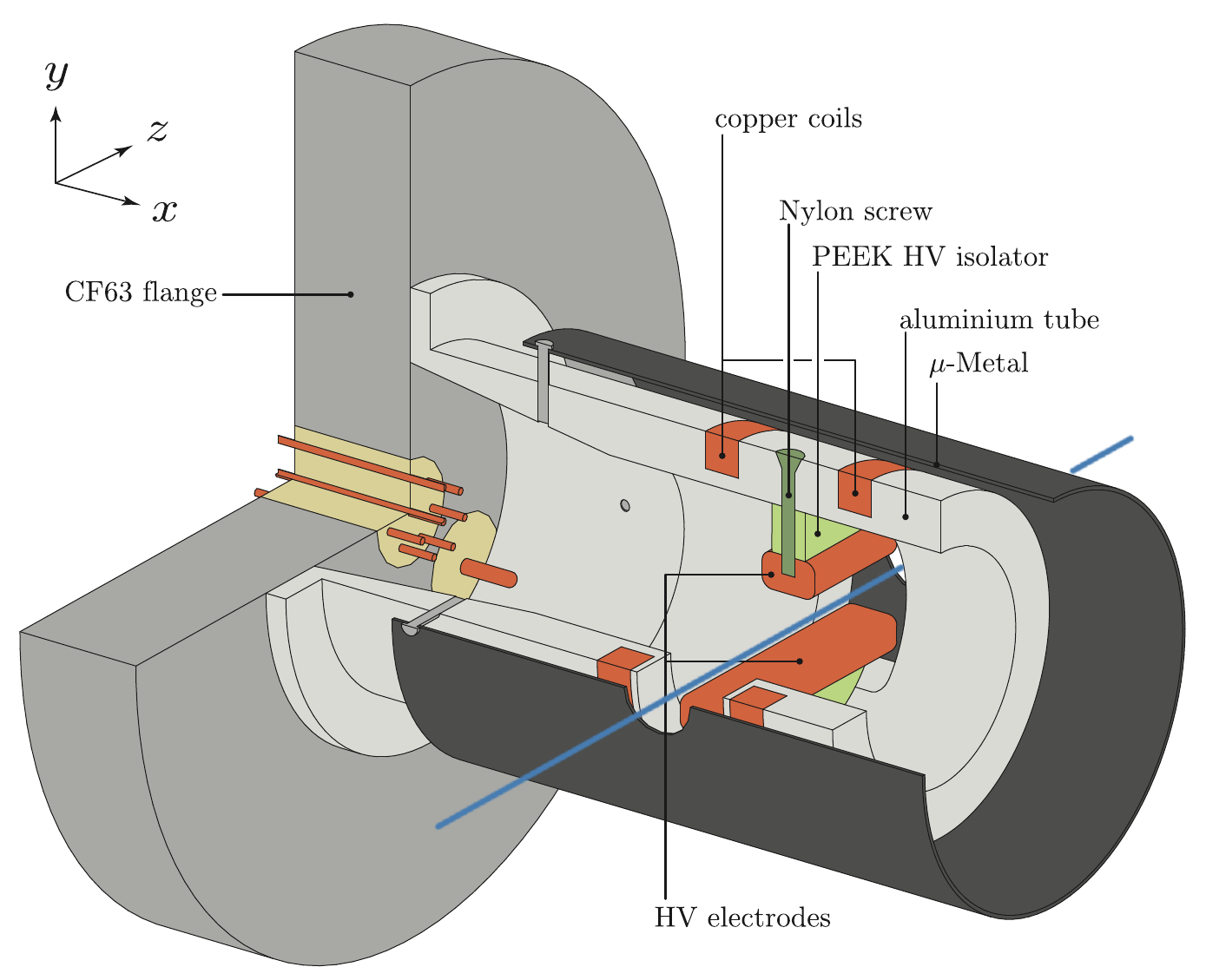}
\caption{\label{Wiendesign}Schematic representation of the Wien filter. The blue line indicates the electron beam. An aluminum cylinder (light grey) is mounted onto the vacuum flange. The green parts represent PEEK insulators. The coils generating the magnetic field $B_{0}~\hat{x}$ and the high-voltage electrodes generating the electric field $E_{0}~\hat{y}$ are indicated in orange.}
\end{figure}

The Wien filter has been designed and built such that it can be inserted into a CF63 vacuum cube. \Fref{Wiendesign} shows a schematic representation of the design. 

\begin{figure}[h]
\centering
\includegraphics[width=10cm]{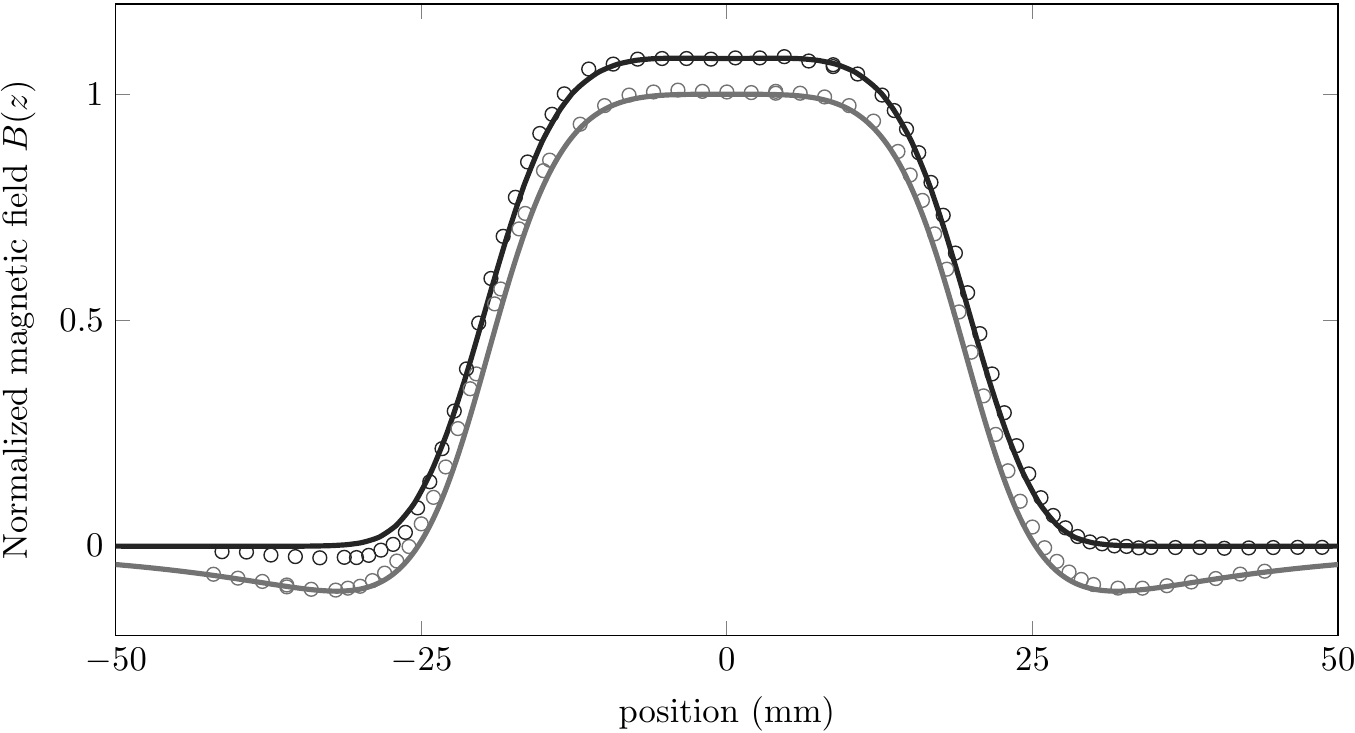}
\caption{\label{WienfilterBfield}The on-axis magnetic field produced by the Wien filter with and without $\mu$-metal shielding. All fields have been normalized with respect to the maximum field without $\mu$-metal. The solid grey line represents the analytically calculated magnetic field without shielding\cite{simpson1829} and the solid black line the simulated magnetic field with $\mu$-metal shielding. The data points represent the measured field profile, both with and without $\mu$-metal.}
\end{figure}

The Wien filter electrostatic field is generated by a pair of capacitor plates. The electrodes are separated by a distance of $7$~mm. Electric field strengths up to $300$~kV/m can be generated. The high-voltage electrodes are suspended by two PEEK insulator wedges. The static magnetic field is produced by a pair of coils in the Helmholtz configuration: the $20$~mm radius of the coil is equal to the distance between the centers. The two coils are wound from $0.6$~mm insulated copper wire with $54$ turns on each coil. The Wien filter is able to produce on-axis magnetic fields up to $5$~mT, which is limited by coil heating.
A $54$~mm diameter, $0.2$~mm thick $\mu$-metal tube provides magnetic shielding. The $\mu$-metal shielding suppresses the magnetic field outside the Wien filter. Without this shielding it is practically impossible to direct the electron beam through the Wien filter. The on-axis magnetic field profiles, both simulated and measured, with and without the $\mu$-metal shielding are depicted in \Fref{WienfilterBfield}. The fields were simulated using the CST\cite{CST} software package and measured using a Hall probe.

\section{Results and discussion}

First we have used the Wien filter to measure the average bunch energy of the electron pulses which agrees with time-of-flight measurements (\Sref{avgbunche}). Hereafter we calibrated the Wien filter (\Sref{seccalib}) and determined the relative energy spread of the electron bunches (\Sref{secEspread}).

All experiments have been performed with $\sim20$ ps electron bunches containing $\sim10^{3}$ electrons produced by a $100~$fs ionization laser pulse\cite{Franssen2017}. The rms spot size of the ionization laser at the position of the MOT was measured to be $\sigma_{\rm ion} = 90\pm10~\mu$m, resulting in an expected rms relative energy spread $\frac{\sigma_{U}}{U}= 0.67\pm0.07\%$ according to \Eref{relenergylasersize}.

\subsection{Average bunch energy}\label{avgbunche}

The electric field in the Wien filter has been determined by measuring voltage across the Wien filter electrodes and the magnetic field by measuring the current running through the coils. The average bunch energy $U$ has been determined using the electric and magnetic field strengths for which no average deflection inside the Wien filter occurred, \Eref{Wiencriterium}. 

\begin{figure*}[htb!]
\centering
\includegraphics[width=10cm]{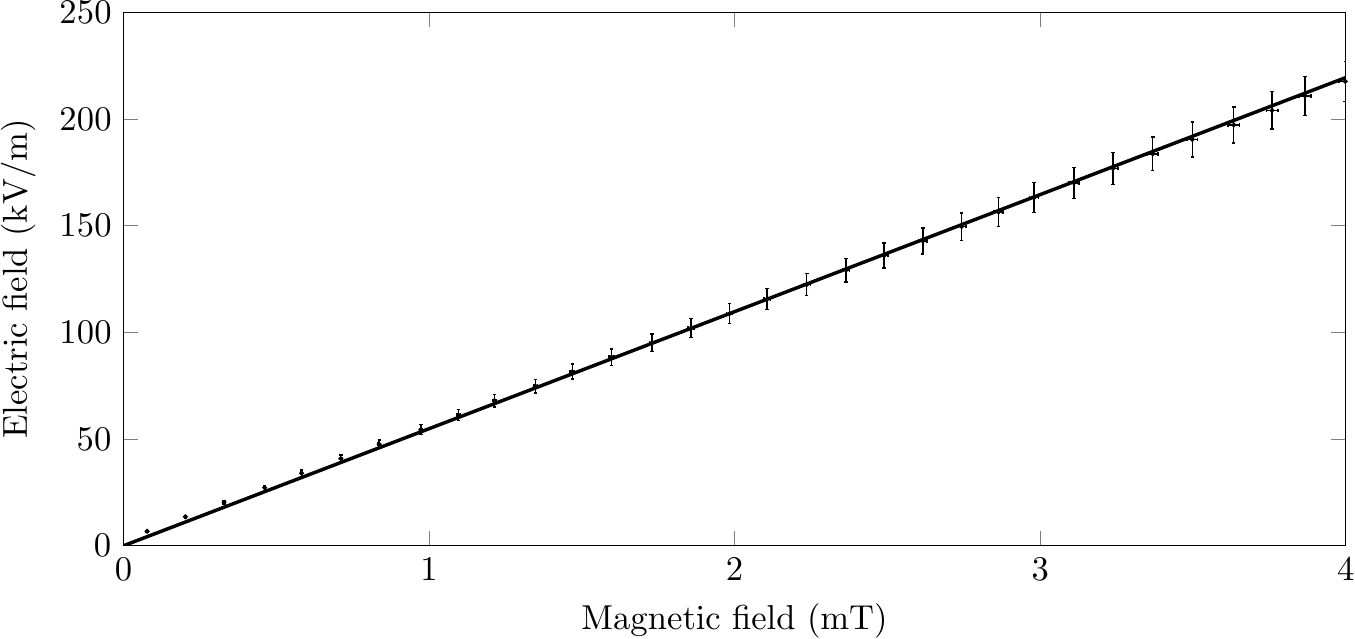}
\caption{\label{WienEB}The electric field a function of the magnetic field, such that the center of the electron bunch is not deflected by the Wien filter. The solid line represents a fit with \Eref{Wiencriterium}.}
\end{figure*}

\Fref{WienEB} shows the electric and magnetic field strengths for which no average deflection occurs together with a linear fit. The electric field is proportional to the voltage on the electrodes and the magnetic field is proportional to the current running through the coils. From the slope of the fit shown in \Fref{WienEB} the average electron beam energy $U$ was determined to be $U=8.8\pm0.2~$keV.

\begin{figure*}[htb!]
\centering
\includegraphics[width=10cm]{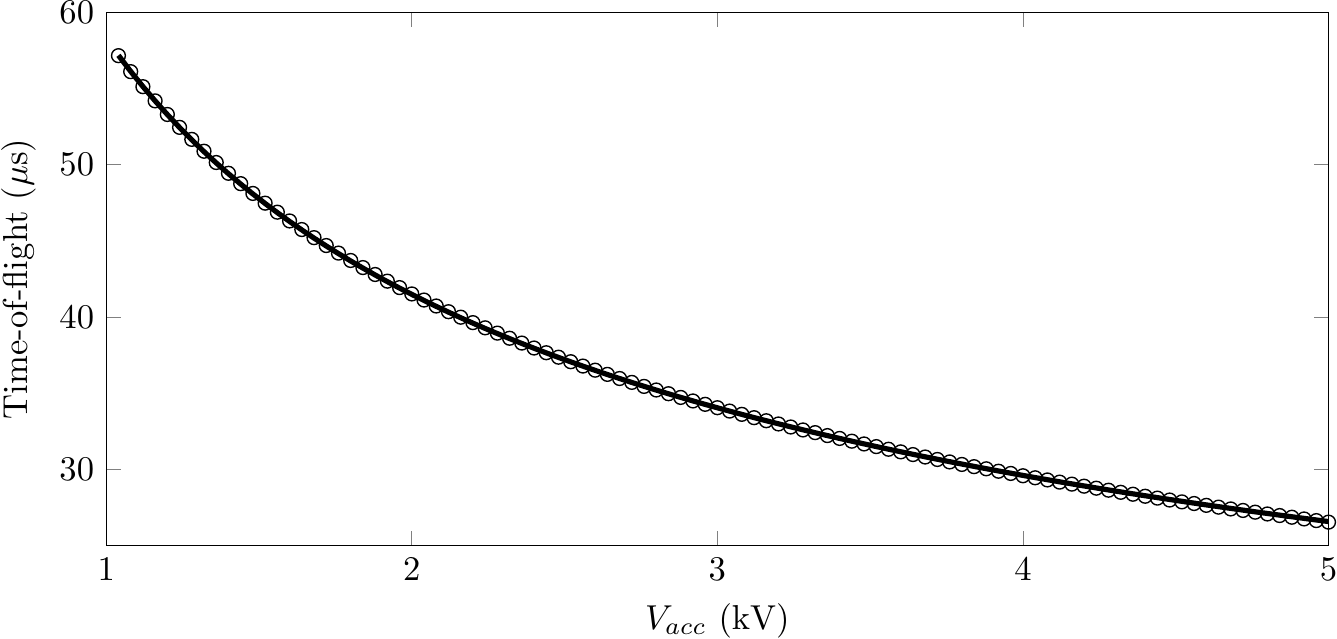}
\caption{\label{TOFscan}The measured ion time-flight as a function of the potential $V_{acc}$ across the accelerating structure together with a fit to our model (solid line). The uncertainty in the data is smaller than the dot size.}
\end{figure*}

The average electron energy was also determined by an ion time-of-fight scan which was done by changing the polarity of the accelerating field. This measurement allows us to determine the position of the ionization volume inside the accelerating structure\cite{Engelen2014,Reijnders2009}. \Fref{TOFscan} shows the ion time-of-flight for various accelerator voltages $V_{acc}$ together with a fit to our model\cite{Engelen2014}. 

From the fit we measure that the average bunch energy $U=(0.50\pm 0.01)\cdot e V_{acc}$. In all the experiments we have used an accelerator potential $V_{acc}=17.5~$kV which results in an average bunch energy of $U=8.7\pm0.2~$keV which agrees well with the bunch energy measured using the Wien Filter.

\subsection{Calibration}\label{seccalib}

\begin{figure*}[htb!]
\centering
\includegraphics[width=10cm]{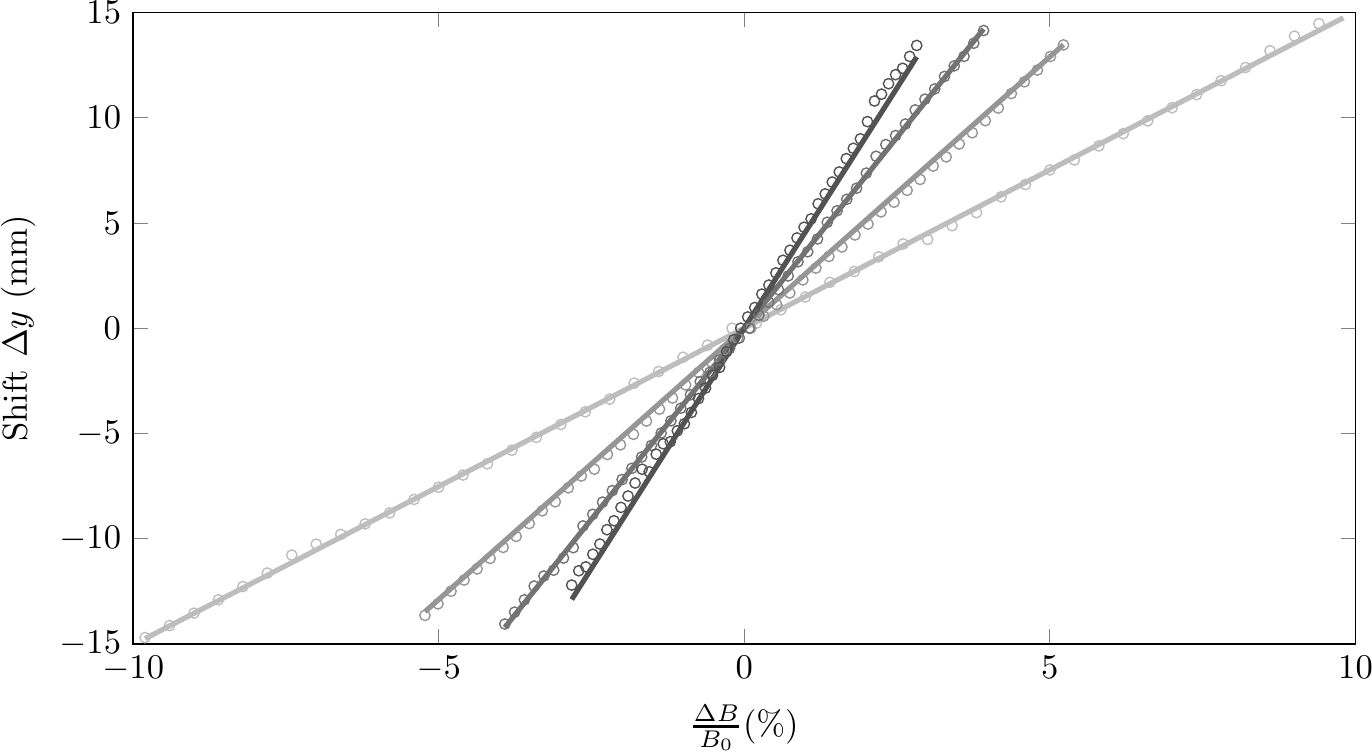}
\caption{The average bunch deflection $\Delta y$ for various relative magnetic field strengths $\frac{\Delta B}{B_{0}} \propto \frac{\Delta I}{I}$ together with linear fits to determine the cyclotron angle $\theta_{c}$. The magnetic field ranges from $B_{0}\approx 1~$mT (light grey), $B_{0}\approx 1.8~$mT, $B_{0}\approx 2.6~$mT, to $B_{0}\approx 3.3~$mT (black).\label{calibrationthetac}}
\end{figure*}

We have calibrated the cyclotron angle $\theta_{c}$ by measuring the average deflection $\Delta y = \frac{\Delta v_{y}}{v_{z}} (d_{det}-d_{wien})$ as a function of magnetic field amplitude at a constant electron energy. \Fref{calibrationthetac} shows the deflection for various relative magnetic field strengths, which scales $\propto \frac{\Delta I}{I}$ with $I$ the current running through the coils. The linear fit through the data in \Fref{calibrationthetac} was used together with \Eref{caldeflec} to determine the cyclotron angle $\theta_{c}$. The magnetic field in \Fref{calibrationthetac} ranges from $B_{0}\approx 1~$mT (light grey), $B_{0}\approx 1.8~$mT, $B_{0}\approx 2.6~$mT, to $B_{0}\approx 3.3~$mT (black).

\begin{figure*}[htb!]
\centering
\includegraphics[width=10cm]{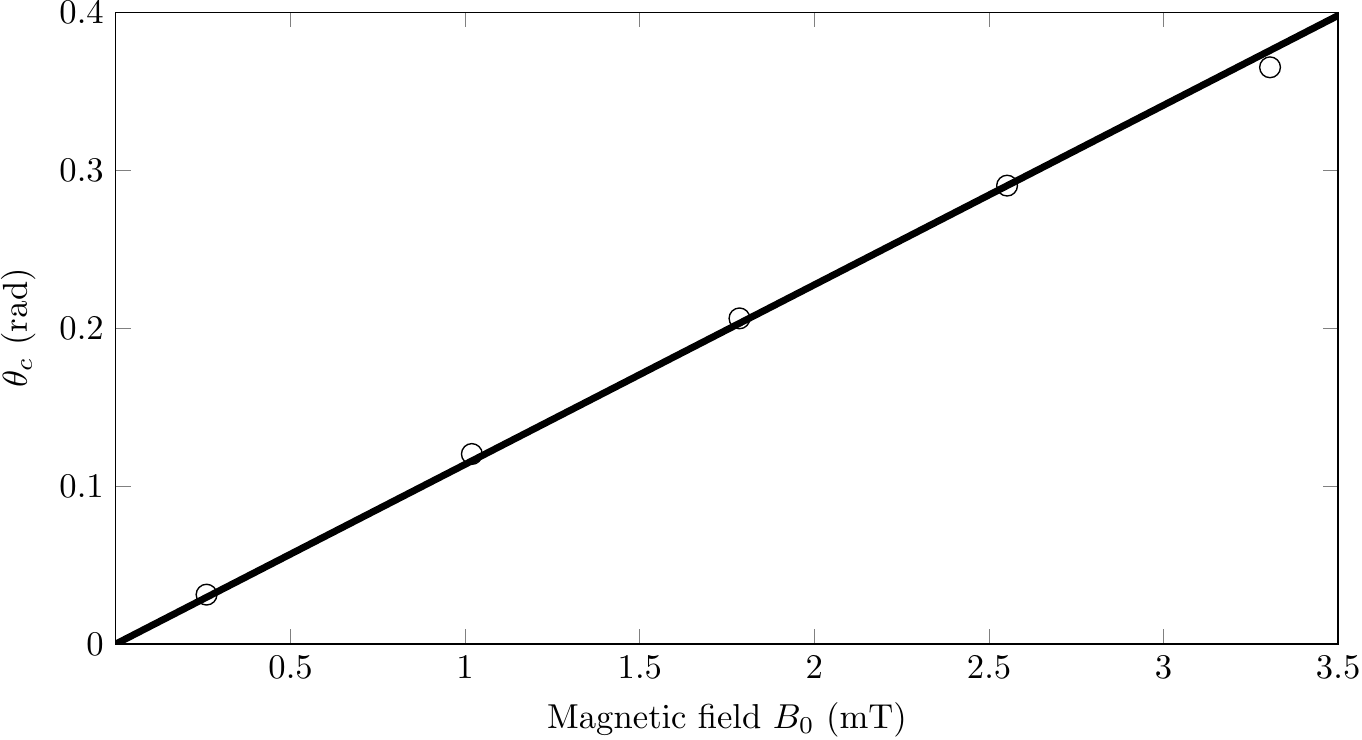}
\caption{The calibrated cyclotron angle (circles) and the theoretical curve (solid) as function of magnetic field strength $B_{0}$ for a fixed electron energy $U=8.74~$keV.  The uncertainty in the data is smaller than the dot size.\label{cyclotroncal}}
\end{figure*}

\Fref{cyclotroncal} shows the calibrated cyclotron angle as a function of magnetic field strength $B_{0}$ together with the theoretical curve which was calculated using $\theta_{c}=\frac{eB_{0}L_{w}}{mv_{z}}$. The figure shows that the Wien filter is working as expected .

\subsection{Energy spread}\label{secEspread}

\begin{figure}[htb!]
\centering
\includegraphics[width=14cm]{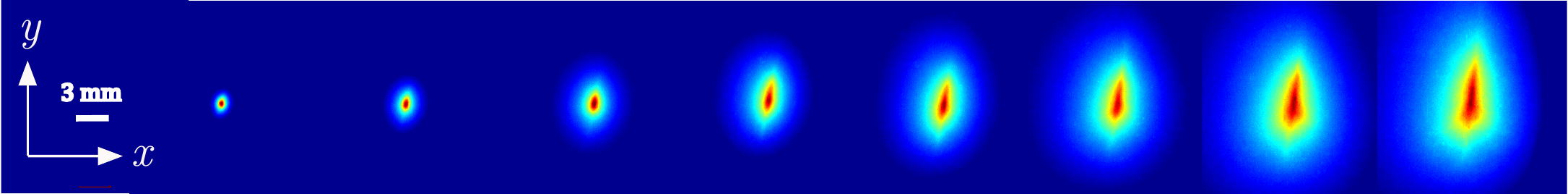}
\caption{\label{Wienstreak}A false color plot of the measured electron distribution as a function of Wien filter cyclotron angle, $\theta_{c}$ from $\theta_{c}=0$ (left) to $\theta_{c}=0.55$ (right). The electron pulse is streaked in the $\hat{y}$ direction.}
\end{figure}

\Fref{Wienstreak} represents a false color plot of the measured spatial electron distribution as a function of cyclotron angle, $\theta_{c}$, from $\theta_{c}=0$ (left) to $\theta_{c}=0.55$ (right). The figure clearly shows that the electron bunch is streaked in the vertical direction. The rms spot size in the $\hat{y}$-direction together with the cyclotron angle $\theta_{c}$ allows us to determine the relative energy spread of the electron bunch.

\begin{figure*}[htb!]
\centering
\includegraphics[width=14cm]{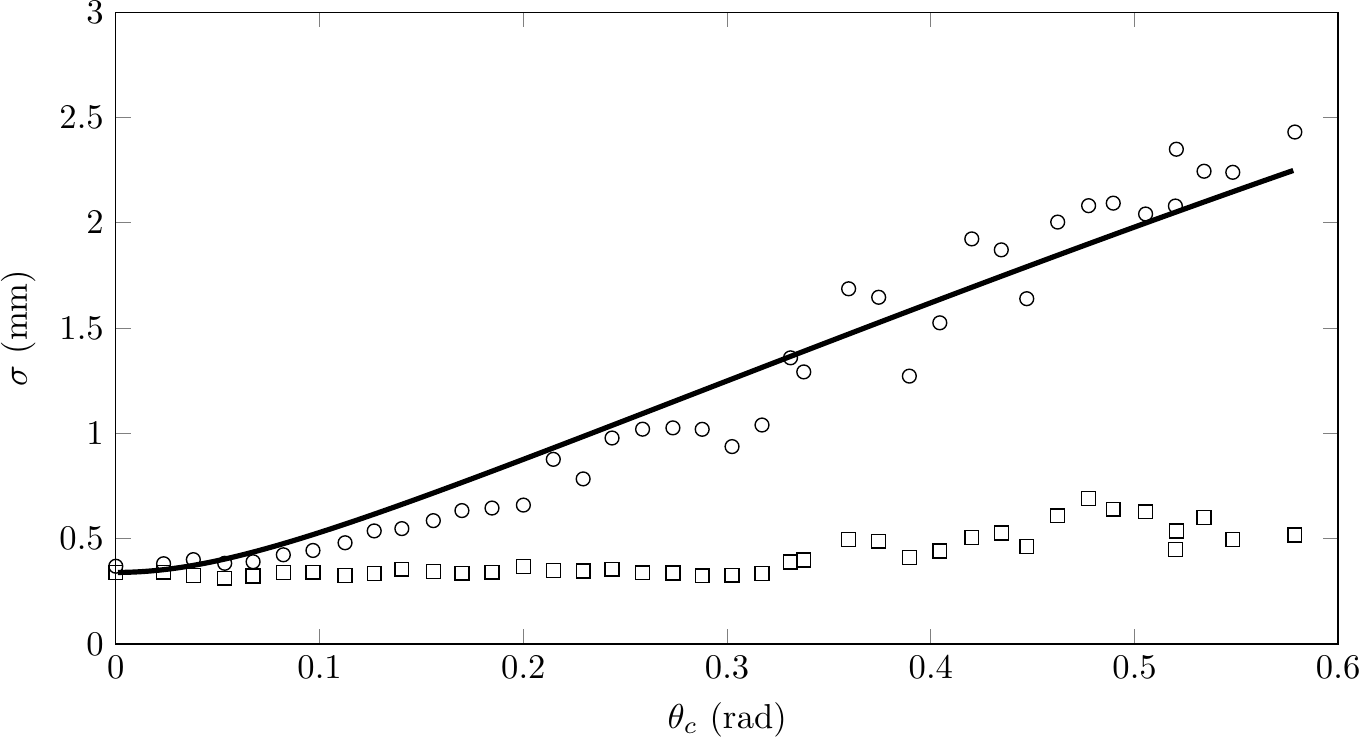}
\caption{\label{rmsstreak} The rms electron spot sizes as a function of cyclotron angle $\theta_{c}$. The circles represent the data parallel to the streak axis and the squares the data perpendicular to the streak axis. The solid line is a fit with \Eref{energyfit}. The uncertainty in the data is smaller than the dot size.}
\end{figure*}

\Fref{rmsstreak} shows the rms size of the electron bunch as a function of the cyclotron angle. The circles indicate the rms size parallel to the streak axis ($\hat{y}$-direction) and the squares the rms size perpendicular to the streak axis ($\hat{x}$-direction). The Wien filter should exert no forces along this direction which is mostly confirmed by the data (squares) presented in \Fref{rmsstreak}. 

The rms relative energy spread was determined by fitting the streak data with \Eref{energyfit} with $\frac{\sigma_{U}}{U}$ the only fitting parameter. This resulted in a rms relative energy spread $\frac{\sigma_{U}}{U}=0.64\pm0.09\%$. This agrees well with the expected value $\frac{\sigma_{U}}{U}=0.67\pm0.07\%$, which is based on the rms size of the ionization laser beam. This is not surprising since the model (\Eref{relenergylasersize}) was already indirectly verified by the energy spread dominated pulse length measurements\cite{Franssen2017} and ion time-of-flight energy spread measurements\cite{Reijnders2009}.

\section{Conclusions and outlook}

The Wien filter has been used to measure the relative energy spread of electron bunches produced by near threshold femtosecond photoionization of a laser cooled and trapped ultracold atomic gas. These are the first measurements of the energy spread of the ultracold electron source. 
The Wien filter has been used to determine the average bunch energy $U=8.8\pm0.2~$keV which agrees with time-of-flight measurements which resulted in $U=8.7\pm0.2~$keV.
The energy spread measurement resulted in $\frac{\sigma_{U}}{U}=0.64 \pm 0.09~\%$, this relative energy spread agrees well with the expected value $\frac{\sigma_{U}}{U}= 0.67\pm0.07\%$ which is based on the rms size of the ionization laser beam. 

Furthermore, the energy spread measurements in combination with pulse length measurements\cite{Franssen2017} will allow us to investigate the full longitudinal phase space distribution of the electron bunches extracted from the ultra cold source. This would make it possible to investigate the longitudinal beam emittance and thus the ultimately achievable temporal resolution in UED experiments for a given energy spread. 


\ack
This research is supported by the Institute of Complex Molecular Systems (ICMS) at Eindhoven University of Technology. Furthermore we thank Eddy Rietman and Harry van Doorn for expert technical assistance.

\section*{References}

\bibliographystyle{iopart-num.bst}

\providecommand{\newblock}{}

\end{document}